\documentstyle[eqsecnum,aps,prd,preprint]{revtex}
\begin{document}
\def\tW{t}
\def\tp{t_{e}}
\def\bx{x}
\def\bk{k}

{\baselineskip0pt
\leftline{\large\sl\vbox to0pt{\hbox{Department of Physics}
               \hbox{Kyoto University}\vss}}
\rightline{\large\rm\vbox to20pt{\hbox{KUNS 1295}
\hbox{\today}\vss}}%
}
\vskip2cm
\begin{center}{\large
    Decoherence on Quantum Tunneling \\
    in the Multi-Dimensional Wave Function Approach}
\end{center}
\begin{center}
 Takahiro~Tanaka \\
{\em Department of Physics,~Kyoto University} \\
{\em Kyoto 606-01,~Japan}
\end{center}
\begin{abstract}
We consider fundamental problems on the understanding
of the tunneling phenomena in the context of the
multi-dimensional wave function.
In this paper, we reconsider the quantum state after
tunneling and extend our previous formalism to the case when
the quantum state before tunneling is in a squeezed state.
Through considering this problem, we reveal that the quantum
decoherence plays a crucial role to allow us of the concise
description of the quantum state after tunneling.

\end{abstract}

\newpage

\section{Introduction}

{}Field theoretical quantum tunneling phenomena such as
false vacuum decay are considered to have played important
roles in the dynamics of the universe in its early stage.
Recently we proposed a possible scenario of the creation
of open universe in one nucleated $O(4)$-symmetric bubble
\cite{Recent}.
Also in the so-called extended inflation scenario \cite{Inf},
the bubble nucleation through the quantum tunneling plays an
important role.

To test these scenarios
by comparing the predictions of the scenarios with the
observed density fluctuations, 
it is required to
investigate the quantum state after tunneling.
For this purpose, we developed a method to investigate the
quantum state after tunneling in
the multi-dimensional wave function approach \cite{TaSaYa},
which was originally investigated in Refs.
\cite{VeGeSa,Vega}.
And we applied it to the problem of the $O(4)$ symmetric
bubble nucleation in Ref. \cite{SaTaYY}.
The quantum state after tunneling
was investigated in slightly different approaches in
Refs. \cite{Rubako,VacVil}

In this paper, we consider a fundamental problem
associated with the quantum tunneling in the multi-dimensional wave
function approach.

First, we review the multi-dimensional
wave function formalism
to construct a WKB tunneling
wave function in the multi-dimensional configuration space.
This WKB wave function naturally defines a WKB time,
which parametrises a sequence of the
configurations corresponding to a classical solution
giving the lowest WKB order description of the wave function.
Usually, we implicitly identify this WKB time with the
external time.
Then, we can give a simple relation between the quantum
state before and after tunneling.
However this identification is not justified {\it a priori},
because the WKB wave function does not describe a statistical
ensemble but it a superposition of the wave packets which denote
the tunneling process occurred at different instants (and locations).

To make the non-triviality of identification explicit,
after the review of our method to determine the
quantum state after tunneling
when the state before tunneling is prepared in a vacuum state
and in an excited state,
we consider an extension to the case where the
state before tunneling is in a squeezed state.

Then we propose a mechanism which allows us to identify
these two different flows of time by applying the idea
used in the discussion
of the quantum decoherence \cite{Deco}.
We consider the situation in which the tunneling degree of freedom
couples to other degrees of freedom whose quantum state is
not measured after tunneling.
We call it the environment.
We consider the the reduced density matrix obtained
by taking a partial trace over these environmental degrees of
freedom.
If the off-diagonal elements of the reduced density matrix
become small and remain so,
the above mentioned identification will be justified
because the nearly diagonal density matrix
can be interpreted as a classical ensemble of different states.
We estimate how effectively this mechanism works.

This paper is organized as follows.
In the previous work, the formalism to determine
the quantum state after tunneling was developed \cite{TaSaYa,Yamamoto}.
However, its derivation was a little bit complicated one.
Therefore, in section 2, we give an intuitive derivation of our
formalism by considering a simple example and we consider an
extension to the case where the
quantum state is in some squeezed state before tunneling.
At the same time, we explain the role of the non-trivial
identification of two different flows of time.
In section 3, we propose a mechanism which allows us to identify
these two different flows of time.
In section 4, brief discussion is in order.

\section{Multi-dimensional Tunneling Wave Function}

In our previous work \cite{TaSaYa},
we developed a method to construct the
multi-dimensional tunneling wave function which describes the
tunneling from the false vacuum ground state.
Although we believe that our previous derivation
was one of the simplest one, it is still complicated because
we considered a rather general situation.
Here we consider a model which is simple
enough for our later discussions but yet contains
essential features of the multi-dimensional quantum tunneling.

We consider a system consisted of one tunneling degree of freedom,
$X$, and $D$ environmental degrees of freedom, $\phi_i$,
coupling to the tunneling
degree of freedom.
The Lagrangian is given by
\begin{equation}
 L={1\over 2}\dot X^2-V(X)+\sum_{i,j=1}^D
   {1\over 2}\delta_{ij} \dot \phi_i \dot \phi_j
   -{1\over 2}m_{ij}^2 (X) \phi_i \phi_j,
\end{equation}
where $X$ and $\phi_i$ are the coordinates for the $D+1$-dimensional
space of dynamical variables.
Eventually, we will be interested in the
extension of this model to field theory,
in which case $\phi_i$ becomes $\phi(x)$ and
the $D+1$-dim. space of dynamical variables becomes
the super space.
We assume the potential $V(X)$ of the form shown in Fig. 1 and
consider the situation in which $X$ is initially trapped in the
false vacuum at $X=X_F$.
This false vacuum decays through quantum tunneling.
When we consider a more realistic situation, $X$
should be interpreted as a collective coordinate.
For simplicity, we have assumed that the environmental or
fluctuation degrees of freedom, $\phi_i$, interact with
the tunneling degree of freedom only through the mass term.
For the later convenience, we define
\begin{equation}
 \omega^2_{ij}:=m_{ij}^2 (X_F),
\end{equation}
and we assume that the coordinate $\phi_i$ is chosen to make
\begin{equation}
 \omega_{ij}=\delta_{ij} \omega_i.
\end{equation}

The Hamiltonian operator in the coordinate representation
is obtained by replacing the conjugate momenta in the
Hamiltonian with the corresponding differential operators as
\begin{equation}
 \hat H=\hat H_X +\hat H_{\phi},
\label{tHam}
\end{equation}
where
\begin{eqnarray}
 \hat H_X & = & -{\hbar^2\over 2}
 \left({\partial^2\over\partial X^2}\right)+V(X),
 \nonumber \\
 \hat H_{\phi} & = & {\hbar^2\over 2}\sum_{i,j=1}^D
 \left(-\delta_{ij}{\partial^2\over\partial \phi_i \partial \phi_j}
 +{1\over 2} m^2_{ij}(X) \phi_i \phi_j\right).
 \label{Hphi}
\end{eqnarray}

\subsection{initially in the {\it quasi-ground-state}}

We construct a wave function which represents the quantum tunneling
phenomena using the WKB approximation.
We call it the {\it quasi-ground-state}\/ wave function.
It is the lowest eigenstate of the Hamiltonian
sufficiently localized in the false vacuum.
When the potential barrier is sufficiently high,
we can approximately define quantum states localized
in the false vacuum.
Let us consider the situation in which
the initial state is set in this {\it quasi-ground-state}\/
localized in the false vacuum.
As the tunneling rate $\Gamma$ is exponentially small,
after a long enough time but not too long compared with the
time scale of the tunneling $\Gamma^{-1}$,
the wave function is expected to become approximately
time independent.
Therefore a {\it quasi-ground-state}\/ wave function
will describe the quantum tunneling from the
{\it quasi-ground-state}\/ in the false vacuum.
To obtain this wave function, we solve the time-independent
Schr\"odinger equation,
\begin{equation}
 \hat H \Psi^0 =E \Psi^0,
 \label{Schroe}
\end{equation}
in the WKB approximation.

If we neglect the environmental degrees of freedom, $\phi$,
the system reduces to that of the one-dimensional
quantum mechanics of a particle.

We impose the WKB ansatz on the wave function,
\begin{equation}
 \Theta=e^{-{1\over \hbar}(W^{(0)}(X)+\hbar W^{(1)}(X)+\cdots)},
\label{Theta}
\end{equation}
 which should solve the time-independent Schr{\"o}dinger equation,
\begin{equation}
 \left[-\hbar^2 {\partial^2\over\partial X^2}+V(X)\right]\Theta(X)
 =E\Theta(X).
\label{tBas}
\end{equation}
We solve this equation to the second lowest order with
respect to $\hbar$.
The energy eigen value $E$ is formally divided into two parts,
$E_0$ and $E_1$, of $O(\hbar^0)$ and $O(\hbar^1)$, respectively.
The equation in the lowest order of $\hbar$ becomes the so-called
Hamilton-Jacobi equation with the energy $E_0\,$,
\begin{equation}
 -{1\over 2}\left({\partial W^{(0)}\over\partial X}\right)^2
 +V(X)=E_0.
  \label{HamJac}
\end{equation}
To obtain a solution of this equation, we introduce a function
$\bar X(\tau)$ which satisfies the relation,
\begin{equation}
 {d\bar X\over d\tau}:={\partial W^{(0)}\over\partial X}.
 \label{defbarx}
\end{equation}
Then the Euclidean equation of motion for $\bar X(\tau)$,
\begin{equation}
 {d^2 X\over d\tau^2}=V'(X),
\end{equation}
is derived from the Eq. (\ref{HamJac}).

We take this solution to start from the false vacuum at
$\tau=-\infty$ with the zero kinetic energy ({\it i.e.},
$E=E_0:=V(X_F)$) and to arrive at the turning point at $\tau=0$
which is the boundary between the classically allowed and forbidden
regions.
It is a half of the so-called instanton solution.
We also call it the dominant escape path (DEP).

Using the definition (\ref{defbarx}), Eq. (\ref{HamJac})
gives
\begin{equation}
 W^{(0)}(\bar X(\tau))=\int_{-\infty}^{\tau}
  d\tau' 2\left(U(\bar X(\tau))-E_0\right) +C',
 \label{W0}
\end{equation}
where $C'$ is a constant.
Therefore, given $\bar X(\tau)$, $W^{(0)}(X)$ can be calculated
using this expression.

In the next order of $\hbar$, Eq. (\ref{Schroe}) gives
\begin{equation}
 -{dW^{(0)}\over dX}{dW^{(1)}\over dX}
 +{1\over 2}{d^2 W^{(0)}\over dX^2}={E_1\over\hbar}.
\end{equation}
As is known well, this equation can be formally integrated to give
\begin{equation}
 W^{(1)}(\bar X(\tau))={1\over 4}\log\left(2(V(\bar X(\tau))
 -E_0\right)-{E_1 \tau\over\hbar}.
 \label{W1}
\end{equation}
Combining Eqs. (\ref{Theta}), (\ref{W0}) and (\ref{W1}),
we obtain the second lowest WKB wave function as
\begin{equation}
 \Theta(\bar X(\tau))={C e^{E_1 \tau/\hbar}\over
 \left(2(V(\bar X(\tau))-E_0\right)^{1/4}}
 \exp \left(-{1\over\hbar}\int_{-\infty}^{\tau}
 d\tau' 2\left(V(\bar X(\tau'))-E_0\right)\right).
 \label{eqTh}
\end{equation}
To see that this has the property of the {\it quasi-ground-state}\/
wave function, we examine the asymptotic behavior of this wave
function near the false vacuum.
There, since locally the potential $V(X)$ may be approximated by
quadratic form as
\begin{equation}
 V(X)=E_0+{1\over 2}\omega_X^2 X^2+\cdots,
\end{equation}
with the definition, $\omega_X^2:={d^2 V\over dX^2}\vert_{X=X_F}$,
we can consider a normalized approximate ground state wave
function in the false vacuum as
\begin{equation}
 \left({\omega_X \over\pi\hbar}\right)^{1/4}
   e^{-{1\over\hbar}\omega_X X^2}.
 \label{harmonic}
\end{equation}
Noting that the DEP is given by
\begin{equation}
 \bar X(\tau)\sim A e^{\omega_X \tau},
\end{equation}
when $\tau$ goes to $-\infty$
where $A$ is a constant,
the requirement that $\Theta(X)$ coincides with (\ref{harmonic})
near the false vacuum determines the unknown two parameters
in (\ref{eqTh}) as
\begin{eqnarray}
 E_1 & = & \hbar\omega_X /2, \nonumber \\
 C & = & (\omega_X^3 A^2/\pi\hbar)^{1/4}.
\end{eqnarray}
The constant $A$ is determined by the condition
${d\bar X(\tau)\over d\tau}=0$
at $\tau=0$, which fixes the origin of time.

Above, we constructed the wave function in the forbidden region.
As is known well, the wave function in the allowed region
is given by its analytic continuation.
Replacing $\tau$ by $i\tW$ and $\bar X(\tau)$ by a solution of
the equation of motion in the Lorentzian time $\tW$,
$\bar X_L(\tW)$, which satisfies
$\bar X_L(\tW)=\bar X(it)$, we obtain
\begin{equation}
 \Theta(\bar X_L (\tW))={C e^{{i\over 2}\omega_X t}\over
 \bigl(2(V(\bar X_L (\tW))-E_0)\bigr)^{1/4}}
 \exp \left({i\over\hbar}\int_{i \infty}^{\tW}
 dt' 2\bigl(V(\bar X_L (t'))-E_0\bigr)\right).
 \label{eqThtwo}
\end{equation}
Here the path of $t$ integration is shown in Fig. 2.
To be more precise, it is necessary to add another term which is
exponentially small in the forbidden region.
However, as it does not change the discussions of the
quantum state after tunneling and the tunneling rate,
we will neglect it in the discussion below.

Next, we consider the system including the environmental
degrees of freedom.
we set an ansatz of the factorised wave function as
\begin{equation}
 \Psi^0 (X,\phi_i)=\Theta(X)\Phi^0 (X,\phi_i).
 \label{finaff}
\end{equation}
Then we find that Eq. (\ref{Schroe}) gives
\begin{equation}
 \left[\hbar{\partial\over\partial\tau}+
   \hat H_{\phi} -E_{1\phi}\right]\Phi^0 (\bar X(\tau),\phi_i)
   =0,
\end{equation}
where $\hat H_{\phi}$ is the Hamiltonian of $\phi_i$
defined in (\ref{Hphi}).
To obtain a solution of this equation, we assume
the Gaussian form of the wave function as
\begin{equation}
 \Phi^0 (\bar X(\tau),\phi_i)
 ={\cal N}(\tau)\exp\left(-{1\over 2\hbar}
 \sum_{i,j=1}^D \Omega_{ij}(\tau)\phi_i \phi_j\right).
\end{equation}
Then a solution of ${\cal N}(\tau)$ and $\Omega_{ij}(\tau)$ are
given by using one matrix, $K_{ij}(\tau)$ as
\begin{eqnarray}
 N(\tau) & = & {\left(\prod_{k=1}^D (\omega_{k}/\pi)\right)^{1/4}
   {1\over\sqrt{\det K_{ij}(\tau)}}}\exp({1\over\hbar}
    E_{1\phi}\tau) \\
 \Omega_{ij}(\tau) & = & \sum_{k=1}^D
     {dK_{ik}\over d\tau}(\tau) K^{-1}_{kj}(\tau),
\end{eqnarray}
and $K_{ij}(\tau)$ satisfies the equation of motion
with respect to $\phi_i$ on the background of $\bar X(\tau)$;
\begin{equation}
 {d^2 K_{ij}\over d\tau^2}(\tau)
 =\sum_{k=1}^D m^2(\bar X(\tau))_{ik} K_{kj}(\tau).
 \label{Keq}
\end{equation}
We need to set an appropriate boundary condition for $K_{ij}(\tau)$
at $\tau=-\infty$ to obtain the {\it quasi-ground-state}\/
wave function.
It is achieved by setting
\begin{equation}
 K_{ij}(\tau)\rightarrow\delta_{ij}\exp(\omega_i \tau)\quad
 \hbox{\rm for}\quad \tau\rightarrow -\infty.
 \label{Kbound}
\end{equation}
In fact, if we choose
\begin{equation}
 E_{1\phi}={\hbar\over 2}\sum_{k=1}^D \omega_{k},
\end{equation}
with this boundary condition the wave function becomes
\begin{equation}
 \Phi^0 (\bar X(\tau),\phi_i)
  \rightarrow\left(\det(\omega/\pi\hbar)\right)
 ^{1/4}\exp\left(-{1\over\hbar}\sum_{k=1}^D\omega_k
 \phi_k^2 \right)\quad
 \hbox{\rm for}\quad \tau\rightarrow -\infty,
\end{equation}
and it coincides with the ground
state wave function approximated by the harmonic potential in
the false vacuum.
Here we comment that $\Omega_{ij}(\tau)$ is symmetric with
respect to the indices, $ij$, because
$\Omega_{ij}(\tau)$ satisfies
\begin{equation}
 \dot \Omega_{ij}(\tau)=m^2_{ij}(\tau)
 -\sum_{k=1}^D \Omega_{ik}(\tau)\Omega_{kj}(\tau),
\end{equation}
which is symmetric,
and so the boundary condition (\ref{Kbound}) is.

The wave function in the allowed region is expected to be given
by its analytic continuation \cite{VacVilC},
\begin{equation}
 \Phi^0 (\bar X_L (\tW),\phi_i)
 ={\cal N}_L (\tW)\exp\left(-{1\over 2\hbar}
 \sum_{i,j=1}^D \Omega_{Lij}\phi_i \phi_j\right),
 \label{envsol}
\end{equation}
where
\begin{eqnarray}
 {\cal N}_L (\tW) & = & {\cal N}(i\tW), \\
 K_{Lij} (\tW) & = & K_{ij}(i\tW), \\
 \Omega_{Lij} & = & -i\sum_{i,j=1}^D K^{-1}_{Lik}(\tW)
    {dK_{Lkj}\over dt}(\tW).
\end{eqnarray}

We note here that, if we define
\begin{equation}
\bar \Phi^0 (\bar X_L (\tW),\phi_i)=
\exp(-iE_{1\phi}\tW/\hbar)\Phi^0 (\bar X_L (\tW),\phi_i),
\end{equation}
factoring out $\phi_i$-independent phase,
$\bar \Phi^0 (\bar X_L (\tW),\phi_i)$ satisfies the
Schr\"odinger equation,
\begin{equation}
  \left[{\hbar\over i}{\partial\over\partial\tW}+
   \hat H_{\phi}\right]\bar\Phi^0 (\bar X_L (\tW),\phi_i) =0,
\end{equation}
with respect to the WKB time
on a given background of $\bar X_L (\tW)$.
We show the quantum state described by
$\bar \Phi^0 (\bar X_L (\tW),\phi_i)$ is a squeezed state.
In order to do so,
let us consider how to represent a squeezed state
in the language of wave function in general.

A squeezed state is a vacuum state in the following sense.
It is naturally described by using a set
of mode functions, $\{u_{ij}(t)\}$.
In the Heisenberg picture, the field operators and its conjugates
are expanded as
\begin{eqnarray}
 \hat\phi_i(t)&=&
 \sum_j \left(u_{ji}(t)A_j +u_{ji}^{*}(t)A^{\dag}_j\right),
\nonumber \\
 \hat p_i(t)&=&
 \sum_j \left(\dot u_{ji}(t)A_j
 +\dot u_{ji}^{*}(t)A^{\dag}_j\right),
\label{modeex}
\end{eqnarray}
by using the mode functions, $\{u_{ij}(\tau)\}$, which solve the
equation of motion,
\begin{equation}
 -{d^2 u_{ij}\over dt^2}(t)
 =\sum_{k=1}^D m^2(\bar X_L(t))_{ik} u_{kj}(t),
\end{equation}
and are
orthonormalised with respect to the Klein-Gordon inner product,
\begin{equation}
(u_{il},u_{jm}):=
      -i\sum_{l,m=1}^D \delta_{lm} \left(u_{il}\dot u^{*}_{jm}
      -\dot u_{il} u^{*}_{jm}
         \right)=\hbar\delta_{ij}.
\label{normKG}
\end{equation}
Then the squeezed state corresponding to $\{u_{ij}(t)\}$ is defined
in the same manner as the usual vacuum state as
\begin{equation}
 A_i \vert O\rangle =0 \quad \hbox{\rm for any}\quad i.
\end{equation}

To move to the Schr\"odinger representation,
we introduce time-dependent annihilation and creation operators
$a_{i}(t)$ and
$a^{\dag}_{i}(t)$, respectively, as
\begin{equation}
 a_{i}(t)
  =U(t)A_{i}U^{\dag}(t),
\quad
 a^{\dag}_{i}(t)
  =U(t)A^{\dag}_{i}U^{\dag}(t),
\label{Schancr}
\end{equation}
and
\begin{equation}
U:=e^{-{i\over\hbar}\int^t dt \hat H_{\phi}},
\end{equation}
where $\hat H_{\phi}$ is a Hamiltonian operator for $\phi_i$
on the background $\bar X_L(t)$.
The Schr\"odinger representations of the field operators
$\hat\phi_{iS}$ and
$\hat p_{iS}$ are given by
\begin{eqnarray}
 \hat\phi_{iS}
  & =& U(t)\hat\phi_i(t)U^{\dag}(t)
\nonumber \\
  & =&\sum_{j=1}^D \left(u_{ji}(t) a_{j}(t)
             +u_{ji}^{*}(t) a^{\dag}_{j}(t)\right),
\nonumber \\
 \hat p_{iS}
  & = &U(t) \hat p_i (t)U^{\dag}(t)
\nonumber \\
  & =& \sum_{j=1}^D \left( \dot u_{ji}(t) a_{j}(t)
       +\dot u_{ji}^{*}(t) a^{\dag}_{ji}(t)\right).
\label{tdadef}
\end{eqnarray}
Using these operators, the Schr\"odinger representation of
the vacuum, {\it i.e.},
$\vert O(t)\rangle_S=U(t) \vert O\rangle$
 is determined by the condition,
\begin{equation}
 a_{i}(t)\vert O(t)\rangle_S=0.
 \label{tdvacdef}
\end{equation}
On the other hand, using the orthonormality of the mode functions,
$a_{i}(t)$ and
$a_{i}^{\dag}(t)$ are expressed as
\begin{eqnarray}
a_{i}(t) &
=&i\sum_{j=1}^D \left( u^{*}_{ij}(t) \hat p_{jS}
 -\dot u^{*}_{ij}(t) \hat\phi_{jS}\right),
 \nonumber \\
 a^{\dag}_{i}(t) &
=&i\sum_{j=1}^D \left(- u_{ij}(t) \hat p_{jS}
  +\dot u_{ij}(t) \hat\phi_{jS}\right).
\label{ancrexp}
\end{eqnarray}
Then, going over to the coordinate representation by
the replacements,
\begin{equation}
\hat p_{iS}\rightarrow -i\hbar
 {\partial\over\partial\phi_i}\,,
\quad
\hat \phi_{iS}\rightarrow \phi_i\,,
\end{equation}
we find from Eq.(\ref{tdvacdef})\ that
\begin{equation}
  \langle\phi_i\vert O(t)\rangle_S={\cal N}(t)
   \exp\left(-{1\over 2\hbar}
  \sum_{i,j=1}^D~\Omega_{ij}(t)\phi_i\phi_j\right),
\label{Schrvac}
\end{equation}
where ${\cal N}(t)$ is a normalization factor and
\begin{equation}
\Omega_{ij}(t)
 ={1\over i}\sum_{k=1}^D \dot u^*_{ik}(t) u^{*-1}_{kj}(t),
\end{equation}
where $u^{-1}_{kj}$ is defined as
\begin{equation}
 \sum_{k=1}^D u_{ik}(t) u^{-1}_{kj}(t)
   =\delta_{ij}.
\end{equation}

Now we are ready to show that
$\bar \Phi^0 (\bar X_L (\tW),\phi_i)$
describes an squeezed state.
{}From Eqs. (\ref{envsol}) and (\ref{Schrvac}), we can read
that $\bar \Phi^0 (\bar X_L (\tW),\phi_i)$ is
a squeezed state represented by the mode functions,
\begin{equation}
 u^*_{ij}(t)=\sum_{k=1}^D c_{ik} {K_{Lkj}(t)\over
    \sqrt{2\hbar\omega_k}},
 \label{Ktou}
\end{equation}
where a constant matrix $c_{ik}$ is chosen to satisfy the
normalization condition (\ref{normKG}).
As a result, the quantum state
after tunneling from the
{\it quasi-ground-state}\/ in the false vacuum
is described by a non-trivial vacuum state
whose mode functions are determined by solving Eq.
(\ref{Keq}) with the boundary condition (\ref{Kbound}).
Their analytic continuations to the Lorentzian region give
the negative frequency functions after the renormalization
given by Eq. (\ref{normKG}).

\subsection{initially in a {\it quasi-excited-state}}

In this subsection, we consider an extension of the
situation discussed in the previous section to the
case in which the quantum state of the environmental degrees
of freedom is in an excited state in the false vacuum,
which we call a {\it quasi-excited-state}.
The arguments presented here are essentially the same as given
in our previous work \cite{EffGra}.
However, to make this paper self-contained, we briefly
repeat them again.

{}Following the procedure taken in Ref. \cite{Yamamoto},
we construct
a set of generalized annihilation and creation operators,
$B_{i}$ and $B^{\dag}_{i}$
\footnote{We used the notation, $B^{\dag}_{i}$, but
$B^{\dag}_{i}$ is not the Hermitian conjugate
operator of $B_{i}$ except at $\tau\rightarrow -\infty$.}
whose action on an eigenstate
of the Hamiltonian produces another eigenstate,
{\it i.e.}, $[\hbar\partial/\partial\tau+\hat H_{\phi},B_i]
=\hbar\omega_{i}B_i$ and $[\hbar\partial/\partial\tau+
\hat H_{\phi},B^{\dag}_{i}]=-\hbar\omega_{i}B^{\dag}_{i}$.
Moreover,
since we look for operators which correspond to the
usual annihilation and creation operators at
the false vacuum origin, we require,
$[\hat H_{\phi},B_i]=\hbar\omega_{i}B_i$ and
$[\hat H_{\phi},B^{\dag}_{i}]=-\hbar\omega_{i}B^{\dag}_{i}$ at
$\tau\rightarrow -\infty$.
Such operators are
\begin{eqnarray}
 \hbar B_i(\tau)&= & e^{-\omega_i\tau}\sum_{j=1}^D
       \left(\sqrt{\hbar\over 2\omega_i} K_{ij}(\tau)
         \hbar{\partial\over\partial\phi_j}
      +\sqrt{\hbar\over 2\omega_i} \dot K_{ij}(\tau)
         \phi_j \right),
 \nonumber \\
 \hbar B^{\dag}_i(\tau)&= &
       e^{\omega_i\tau}\sum_{j=1}^D
       \left(-\sqrt{\hbar\over 2\omega_i} Q_{ij}(\tau)
       \hbar{\partial\over\partial\phi_j}
       -\sqrt{\hbar\over 2\omega_i}\dot Q_{ij}(\tau)
       \phi_j\right),
 \label{Bdag}
\end{eqnarray}
where $Q_{ij}$ is assumed to satisfy the same equation
as Eq.(\ref{Keq})\ for $K_{ij}$ but with the
opposite boundary condition as
\begin{equation}
 Q_{ij}(\tau) \rightarrow \delta_{ij}
      e^{-\omega_i \tau} \quad{\rm for}~
         \tau\rightarrow-\infty\,.
\label{InitK}
\end{equation}
In fact, these operators reduce to the ordinary
annihilation and creation operators in the Heisenberg
representation in the false vacuum like
\begin{eqnarray}
 \hbar B_i(\tau)& \rightarrow &
        \sqrt{\hbar\over 2\omega_i}\left(
        \hbar{\partial\over\partial\phi_i}
        +\omega_i \phi_i\right):=\hbar A_{Fi}, \nonumber \\
 \hbar B^{\dag}_i(\tau)& \rightarrow &
       -\sqrt{\hbar\over 2\omega_i}\left(
       \hbar{\partial\over\partial\phi_i}
       -\omega_i \phi_i\right):=\hbar A^{\dag}_{Fi},
    \quad (\tau\rightarrow -\infty).
 \label{Fidef}
\end{eqnarray}
Therefore a {\it quasi-excited-state} wave function with
respect to
the environmental degrees of freedom can be obtained by
operating these creation operators, $B^{\dag}_i(\tau)$,
to the {{\it quasi-ground-state}\/} wave function as
\begin{equation}
 \Psi^{n_1,n_2,\cdots,n_D}(\bar X (\tau),\phi_i) =
 \prod_{i=1}^D
 \{B^{\dag}_i(\tau)\}^{n_i}\Psi^0(\bar X (\tau),\phi_i).
 \label{excited}
\end{equation}
The energy eigen value of this wave function is
\begin{equation}
 E_{n_1,n_2,\cdots,n_D}=E_0+E_{1}+\hbar\sum_{i=1}^D
    \left(n_i+{1\over 2}\right) \omega_i.
\end{equation}

As in the previous case,
factoring out the $\phi_i$ independent part in
$\Psi^{n_1,n_2,\cdots,n_D}(\bar X (\tau),\phi_i)$,
we can extract
$\bar \Phi^{n_1,n_2,\cdots,n_D}(\bar X_L (\tW),\phi_i)$
which satisfies
the Schr\"odinger equation on the background, $\bar X_L(t)$,
with respect to the WKB time.
Introducing
\begin{eqnarray}
b^{\dag}_i(\tW) &:= &e^{-i\omega_i t}B^{\dag}_i(\tW),
\nonumber \\
  & = & \sum_{j=1}^D
       \left(-\sqrt{\hbar\over 2\omega_i} Q_{ij}(\tW)
       \hbar{\partial\over\partial\phi_j}
       +i\sqrt{\hbar\over 2\omega_i}\dot Q_{ij}(\tW)
       \phi_j\right),
  \label{bdag}
\end{eqnarray}
it is explicitly written as
\begin{equation}
 \bar \Phi^{n_1,n_2,\cdots,n_D}(\bar X_L (\tW),\phi_i)
  = \prod_{i=1}^D \{b^{\dag}_i(\tW)\}^{n_i}
 \bar\Phi^0(\bar X_L (\tW),\phi_i).
\end{equation}

The quantum state described by
$\bar \Phi^{n_1,n_2,\cdots,n_D}(\bar X_L (\tW),\phi_i)$
can be understood in a more transparent way
in the Heisenberg picture.
Using the mode functions defined in Eq. (\ref{Ktou}),
the mode functions $Q_{Lij}(\tW)$ are expanded
in the Lorentzian region as
\begin{equation}
 {1\over \sqrt{2\hbar\omega_i}}
  Q_{Lij}(t)=\sum_{k=1}^D r_{ik} u_{kj}(t)+s_{ik} u^{*}_{kj}(t),
 \label{QExp}
\end{equation}
where $r_{kj}$ and $s_{kj}$ are constant matrices.
Then, comparing Eqs.~(\ref{ancrexp}), (\ref{bdag})
and (\ref{QExp}),
we obtain the representation of $b_i^{\dag}(t)$ in
terms of the annihilation and creation operators
$a^{\dag}_{j}(t)$ and $a_{j}(t)$ associated
with $u_{ij}(t)$ like
\begin{equation}
  b^{\dag}_i(t)
 =\sum_{k=1}^D \left(
 r_{ij} a^{\dag}_{j}(t)+s_{ij} a_{j}(t)\right)
 :=b^{\dag}_i(\{a_{j}(t),a^{\dag}_{j}(t)\}).
\end{equation}
We find that $b^{\dag}_i(t)$ is a linear combination of
$a_{j}(t)$ and $a^{\dag}_{j}(t)$.
Therefore the quantum state after tunneling is not represented
as a simple excited state on the squeezed vacuum corresponding
to the set of mode functions, $u_{ij}(t)$,
but a superposition of different excited states
which is obtained by the different number of operations of
the creation (and annihilation) operators associated with
these mode functions.
In the Heisenberg representation, this wave function is written as
\begin{eqnarray}
 \vert n_1,n_2,\cdots,n_D\rangle & \propto &
  U\prod_{i=1}^D
 \left\{b_i^{\dag}(\{a_i (t),a^{\dag}_{i} (t)\})
 \right\}^{n_i}U^{\dag} U\bar\Phi^0(\bar X_L (\tW),\phi_i)
 \nonumber \\
 & = & \prod_{i=1}^D
 \left\{b_i^{\dag}(\{A_i ,A^{\dag}_{i}\})
 \right\}^{n_i}\vert O\rangle,
\end{eqnarray}
where $A_i$  and $A^{\dag}_i$ are the same ones defined in the
previous subsection.

\subsection{initially in a {\it quasi-squeezed-state}}

In this subsection, we consider the case in which the quantum state
of the environmental degrees of freedom is in some
squeezed state in the false vacuum.

A {\it quasi-squeezed-state}
is determined by a set of mode functions in false vacuum,
$\{\bar u_{Fij}(t)\}$.
These mode functions are
expanded by the false vacuum mode functions, $u_{Fij}(t):=
\sqrt{\hbar/2\omega_i} \delta_{ij} e^{-i\omega_i t}$, as
\begin{equation}
\bar u_{Fij}(t)=\sum_{k=1}^D \alpha_{ik} u_{Fkj}(t)
   +\beta_{ik} u^{*}_{Fkj}(t),
 \label{sqvac}
\end{equation}
where $\alpha_{ik}$ and $\beta_{ik}$ are so-called
Bogoliubov coefficients.
The annihilation and creation operators associated with $u_{Fij}(t)$
are, respectively, $A_{Fi}$ and $A^{\dag}_{Fi}$ given in
Eq. (\ref{Fidef}).

This squeezed state is a superposition of
different excited states on the {\it quasi-ground-state} and it can
be represented concisely in the Heisenberg picture as
\begin{equation}
 \vert \alpha,\beta\rangle_F
 = {\cal N}\exp\left({1\over 2}
 \sum_{i,j=1}^D (\alpha^{*-1}\beta^{*})_{ij}
 A_{Fi}^{\dag} A_{Fj}^{\dag}\right)\vert O \rangle_F,
 \label{squeezed}
\end{equation}
where ${\cal N}$ is a normalization constant.
Actually, noting
that the matrix $(\alpha^{*-1}\beta^{*})_{ij}$
is symmetric, which is proved by using the relation
generally satisfied by
Bogoliubov coefficients; $\displaystyle \sum_{k=1}^D \left(
\alpha_{ik} \beta_{jk}-\beta_{ik} \alpha_{jk}\right)=0$,
we can show that by the action of
annihilation operators associated with
$\{\bar u_{Fij}(t)\}$,
\begin{equation}
 \bar A_{Fi}=\sum_{j=1}^D
    \alpha_{ij}^{*}A_{Fj}-\beta^{*}_{ij}A^{\dag}_{Fj},
\end{equation}
$\vert\alpha,\beta\rangle_F$ is annihilated.

When we translate it into the language of the wave function,
we need to be aware that the squeezed state is
not an eigen state of the Hamiltonian but a superposition
of many excited states with different energy.
So far, as we considered only an energy eigen state,
a time independent wave function was sufficient.
But it is not the case for a {\it quasi-squeezed state}
wave function.
Therefore we must consider a time dependent wave function
introducing the external time $\tp$ which is different from
the WKB time $\tW$.
(We may need to mention that there is no WKB time
in the false vacuum.)
Then the wave function corresponding to
$\vert \alpha,\beta\rangle_F $
is represented as
\begin{eqnarray}
 \Psi^{\{\alpha,\beta\}}_F (X_F,\phi_i;\tp) & = &
  \langle\phi_i,X\vert e^{-i\hat H \tp/\hbar}
   \vert\alpha ,\beta\rangle_F
   \nonumber \\
  & = & {\cal N}\exp\left({1\over 2}\sum_{i,j=1}^D
  (\alpha^{*-1}\beta^{*})_{ij}
  A^{\dag}_{Fi} A^{\dag}_{Fj}
   e^{-i(\omega_i+\omega_j)\tp}\right)\Psi^0(X_F ,\phi_i),
\end{eqnarray}
in the false vacuum.
So the wave function is represented as a superposition of
excited state wave functions there.
Since the tunneling wave function for
each excited state is already
constructed in the previous subsection,
the wave function with this asymptotic behaviour
can be obtained by the similar superposition of the excited
state wave functions $\Psi^{n_1,n_2,\cdots,n_D}$
defined in Eq (\ref{excited}).
Therefore the wave
function after tunneling can be described by
\begin{eqnarray}
  \Psi^{\{\alpha,\beta\}} (\bar X_L &(\tW)&,\phi_i;\tp)
    =  {\cal N}\exp\left({1\over 2}\sum_{i,j=1}^D
  (\alpha^{*-1}\beta^{*})_{ij}
  B^{\dag}_{i}(\tW) B^{\dag}_{j}(\tW)
   e^{-i(\omega_i+\omega_j)\tp}\right)
   \Psi^0(\bar X_L (\tW),\phi_i)
  \nonumber \\
   & = & {\cal N}\exp\left({1\over 2}\sum_{i,j=1}^D
  (\alpha^{*-1}\beta^{*})_{ij}
  e^{-i\omega_i (\tp-t)}b^{\dag}_{i}(\tW)
  e^{-i\omega_j (\tp-t)}b^{\dag}_{j}(\tW)
  \right)\Psi^0(\bar X_L (\tW),\phi_i) .
\end{eqnarray}
This state is specified by the following annihilation operators,
\begin{equation}
 \bar B_i(t;\tp)
   := \alpha^{*}_{ij}e^{-i\omega_j (\tW-\tp)}b_j (\tW)
    -\beta^{*}_{ij}e^{i\omega_j (\tW-\tp)}b_j^{\dag} (\tW).
\end{equation}
It is easy to see that $\bar B_i(t;\tp)\Psi^{\{\alpha,\beta\}}
(\bar X_L (\tW),\phi_i;\tp)=0$ holds.

Here we find that a concise
statement on the quantum state after tunneling can be made,
provided the flows of these two different notions
of time are identical.
If it is the case, we may set $t-\tp=\delta=const.$ and
$\bar B_i$ becomes
\begin{equation}
 \hbar \bar B_i(t)
    = \sum_{j,k=1}^D \sqrt{\hbar\over 2\omega_j}
     \bar K_{Ljk} (t)
     \hbar{\partial\over\partial\phi_k}
     -i\sqrt{\hbar\over 2\omega_j}
     {d\bar K_{Ljk}\over dt} (t)\phi_k,
 \label{Bbar}
\end{equation}
where
\begin{equation}
     \bar K_{Ljk} (t)=
     \alpha^{*}_{ij} e^{-i\omega_j \delta}
     K_{Ljk} (t) +\beta^{*}_{ij}
     e^{i\omega_j \delta} Q_{Ljk} (t).
\end{equation}
Therefore the quantum state after tunneling becomes a squeezed
state with the negative frequency functions,
\begin{equation}
 \bar u^{*}_{ij} (t)=\sum_{k=1}^D \bar c_{ik}\bar K_{Lkj} (t),
\end{equation}
where $\bar c_{ik}$ is a constant matrix chosen to satisfy
the orthonormality of $\bar u^{*}_{ij}$.
Moreover, $\bar K_{Lkj} (t)$ solves the
equation of motion along the DEP shown in
Fig.~2 with the initial condition given by
\begin{eqnarray}
\bar u^{*}_{Fij}(t;\delta) & = & \sum_{k=1}^D \alpha_{ik}
     e^{-i\omega_k \delta} u_{Fkj}(t)
   +\beta_{ik} e^{i\omega_k \delta} u^{*}_{Fkj}(t)
 \nonumber \\
    & = & \bar u^{*}_{Fij}(t-\delta).
\end{eqnarray}
So we conclude that the quantum state after tunneling
is determined by the mode functions which solve the equation of
motion along the tunneling background with the boundary condition
that they coincide with the negative frequency functions in the
false vacuum as in the case of tunneling from the
{\it quasi-ground-state}\/.

The above statement is very simple, but the identification
of two different flows of time is non-trivial.
On this point we discuss in the next section.

\section{Decoherence and identification of two different flows of
time}

In the previous section, we pointed out that the
identification of two different flows of time, {\it i.e.},
the WKB time and the external time,
plays an important role in the interpretation
of the quantum state after tunneling.
Hence it is important if we can justify this identification.
One may say that the lowest WKB description gives a
classical trajectory already and we do not have to distinguish
these two flows of time from the beginning.
Judging from ordinary experiences, one may feel
this statement is correct.
But it should be justified more rigorously.

How the classical behaviors of the system appear
is a very interesting
topic in the quantum cosmology 
as well as in the theory of the quantum measurement. 
As long as the evolution of a system governed by a Hamiltonian is
considered, it must be unitary.
Therefore if the quantum state is prepared in a pure state, it remains
so forever, even though it is a superposition of
macroscopically different states.
It seems to contradict with our ordinary experiences.
The most conservative way of thinking to understand this paradox
is given in the context of quantum decoherence in an
open system \cite{Deco}.
There, the total system is divided into two parts,
{\it i.e.},system and environment.
In reality, there are many unseen degrees of freedom,
which are called environmental degrees of freedom here.
When we evaluate the expectation value of the operator
belonging to the system, we do not have to know the density
matrix of the total system but the reduced density matrix is enough.
The reduced density matrix is given by taking a partial trace of
the density matrix with respect to the environmental degrees of freedom.
The important point is that this reduced density matrix does not
necessarily remain in a pure state any longer
even if it is initially so.
Generally, It evolves into a mixed state.
A mixed state may be understood as
a statistical ensemble of different quantum states,
which we call sectors.
When each sector has a rather sharp peak in the
probability distribution of the operators
and the evolution of
each sector is approximately independent of each other,
{\it i.e.}, when the quantum coherence between
different sectors is lost,
then the system is recognized to become classical.

Here, we do not discuss fundamental issues of the quantum
decoherence and the classicallity.
Instead, we follow the standard discussion about decoherence
and apply it to the tunneling system.
Following the usual strategy \cite{strate},
we calculate the reduced density matrix
and estimate its off-diagonal elements.
In the present case, the system is composed of
the tunneling degree of freedom,
$X$, and a part of environmental degrees of freedom whose quantum
state after tunneling we are interested in,
$\phi_i~ (i=D'+1,\cdots D)$,
and the remaining environmental degrees of freedom,
 $\phi_i~(i=1,\cdots D')$, which we do not measure.
As was shown in the previous section, a simple representation of the
quantum state after tunneling is allowed only when we have a
good reason to identify the flow of the WKB time with
that of the external time.
The WKB wave function is considered as a superposition of wave
packets which tunnels through the barrier at different instances.
These wave packets are considered as sectors here and
they are labeled by the values of $\delta$.
In each sector labeled by $\delta$,
the flow of the WKB time and that of
the external time can be
identified as $t-\tp\sim\delta$
within the precision of the broadness of the wave packet.
Therefore, we examine below to what extent
the coherence between the states corresponding to
different $\delta$ is lost.
This is equivalent to examine the degree of decoherence
between the states of different WKB time at a given external
time $\tp$.
If the correlation between them is lost, we can say that
the identification of two different flows of time is allowed.
In practice, we evaluate how small the off-diagonal elements
of the reduced density matrix become
when it is represented in the coordinate basis.

We are interested in the case initially in a
{\it quasi-squeezed-state} but the decoherence between
different sectors also occurs in
the case initially in the {\it quasi-ground-state}\/.
Therefore, for simplicity, the latter case is considered first,
and the modification to the former case is examined later.

The total density matrix for the {\it quasi-ground-state}
is given by a product of the wave
function obtained in the previous section as
\begin{equation}
 \rho(\bar X_L (\tW),\phi_i;\bar X_L (\tW'),\phi'_i;\tp):=
  \Psi^{0} (\bar X_L (\tW),\phi_i)
  \Psi^{0*}(\bar X_L (\tW'),\phi'_i).
\end{equation}
Since the density matrix becomes time independent,
we omit $\tp$ for the notational simplicity in the following discussion.
The reduced density matrix is given by taking a partial trace with respect
to the environmental degrees of freedom like
\begin{equation}
 \tilde\rho(\bar X_L (\tW);\bar X_L (\tW'))
   :=\prod_{i=1}^{D'} \left\{\int_{-\infty}^{\infty}
   d\phi_i\right\} \rho(\bar X_L (\tW),\phi_i;\bar X_L (\tW'),\phi'_i).
 \label{tilderho}
\end{equation}
When  each $\phi$ decouples from each other, or equivalently,
when
\begin{equation}
 m^2_{ij}(X)=m^2_i(X) \delta_{ij},
\end{equation}
we can deal with each $\phi$ separately.
To avoid the unnecessary complexity, let us further
assume that the mass becomes constant after tunneling as
\begin{equation}
 m^2_i (\bar X_L (\tW))=m^2_{Ti}.
\end{equation}
Then the wave function becomes
\begin{equation}
 \Psi^{0} (\bar X_L (\tW),\phi_i)
 =\Theta(\bar X_L (\tW))
 \prod_{i=1}^{D'} \Phi^{0}_i(\bar X_L (\tW),\phi_i),
\end{equation}
and
\begin{equation}
 \Phi^{0}_i(\bar X_L (\tW),\phi_i):=
 \tilde{\cal N}(t)\tilde \Phi^{0}_i(\bar X_L (\tW),\phi_i)
   =\tilde{\cal N}(t)
   \left(\Re(\Omega_i (\tW))\over\pi\hbar\right)^{1/4}
   \exp\left(-{\Omega_i(\tW)\over 2\hbar}\phi_i^2\right),
\end{equation}
where
\begin{eqnarray}
 \tilde{\cal N}(t) & = & {\cal N}(t)
   /\left(\Re(\Omega_i (\tW))\over\pi\hbar\right)^{1/4},
 \nonumber \\
 \Omega_i (\tW) & = & {dK_{Li}(\tW)\over id\tW}K_{Li}^{-1}(\tW),
\end{eqnarray}
and $K_{i}(\tau)$ satisfy the equation
$\ddot K_i =m^2_i(\bar X(\tau)) K_i$
in the Euclidean region.
Generally, up to overall normalization, $K_{Li}(\tW)$ is specified by
two real parameters $\gamma_i$ and $\varphi_i$ like
\begin{equation}
 K_{Li}(\tW)={\cal C}_i
  \left( e^{-im_{Ti}\tW} + e^{-2\gamma_i+2i\varphi_i}
  e^{im_{Ti}\tW}\right).
\end{equation}

Then $\tilde\rho(\bar X_L (\tW);\bar X_L (\tW'))$ is expressed as
\begin{equation}
 \tilde\rho(\bar X_L (\tW);\bar X_L (\tW'))
 =\Theta(\bar X_L (\tW)) \Theta^{*}(\bar X_L (\tW'))
  \prod_{i=1}^{D'} \left\{
   \tilde{\cal N}(t) \tilde{\cal N}^{*}(t')\right\}
  \prod_{i=1}^{D'} {\cal R}_i (t,t'),
\end{equation}
where
\begin{eqnarray}
 {\cal R}_i(t,t') &:=&
 \left({\Re(\Omega_i (\tW))\Re(\Omega_i (\tW'))
 \over\pi^2\hbar}\right)^{1/4}
 \int_{-\infty}^{\infty} d\phi_i
 \exp\left(-{\Omega_i (\tW)\over 2\hbar}\phi_i^2\right)
 \exp\left(-{\Omega_i^{*}(\tW')\over 2\hbar}\phi_i^2\right)
 \nonumber \\
 & = & \left(2\sqrt{\Re(\Omega_i (\tW)\Omega_i (\tW'))}
 \over\left\vert\Omega_i (\tW)
 +\Omega_i^{*}(\tW')\right\vert\right)^{1/2}.
 \label{Omegai}
\end{eqnarray}
The factor
$\displaystyle \prod_{i=1}^{D'} {\cal R}_i(t,t')$ gives the
relative amplitude of the off diagonal elements of
the density matrix to the diagonal elements.
{}From this expression, we can show that ${\cal R}_i(t,t')\leq 1$
and the equality holds only when $\Omega_i (\tW)=\Omega_i (\tW')$.
Especially, ${\cal R}_i(t,t)=1$.

When the difference $\Delta\Omega_i:=\Omega_i (\tW)-\Omega_i
(\tW')$ is small,
the above expression reduces to the following simple one,
\begin{equation}
 {\cal R}_i (t,t') =1-{1\over 16\{\Re(\Omega_i(\tW))\}^2}
   \vert\Delta\Omega_i\vert^2
 +O\left(\left({\Delta\Omega_i\over\Re(\Omega_i(\tW))}\right)^3\right).
\end{equation}
Moreover, using
\begin{equation}
 {d\Omega_i (\tW)\over id\tW}=m^2_{Ti}-\Omega_i^2 (\tW),
\end{equation}
we can show that
\begin{equation}
 \vert\Delta\Omega_i\vert^2=
  \left\vert{d\Omega_i (\tW)\over id\tW}\right\vert^2 (t-t')^2
 ={4m^4_{Ti}(t-t')^2 \over
   (\cosh 2\gamma_i +\cos 2(m_{Ti}\tW+\varphi_i))^2},
\end{equation}
and
\begin{equation}
 \Re(\Omega_i(\tW))=m_{Ti}{\sinh 2\gamma_i\over\cosh 2\gamma_i +
 \cos 2(m_{Ti}\tW+\varphi_i)}.
 \label{ReOmega}
\end{equation}
Then we obtain
\begin{equation}
 {\cal R}_i(t,t')=1-{m^2_{Ti}\over 4\sinh 2\gamma_i}(\tW-\tW')^2
 +O\left(\left({\Delta\Omega_i\over
 \Re(\Omega_i(\tW))}\right)^3 \right).
 \label{vacR}
\end{equation}
We find that, in the present case, the dependence of
${\cal R}_i(t,t')$
on $\tW$ and $\tW'$ becomes very simple.
Also, from this expression, ${\cal R}_i(t,t')$ are found to be
independent of the phase $\varphi_i$.
The dependence on $\gamma_i$ is easy to be understood.
Since a small excitation corresponds to large value of
$\gamma_i$ and a large excitation corresponds to $\gamma_i\sim 0$,
we can say that the coherence factor becomes small
when the environment is highly excited and, on the other hand, it
becomes close to unity when the environment
remains nearly in the vacuum state.

Next we consider the case
initially in the squeezed state.
{}From Eq.~(\ref{Bbar}), we can extract the $\phi$ dependence
of the wave function as
\begin{equation}
 \tilde \Phi^{\alpha\beta}_i(\bar X_L (\tW),\phi_i;\tp)=
   \left(\Re(\Omega_i (\tW;\tp))\over\pi\hbar\right)^{1/4}
   \exp\left(-{\Omega_i(\tW;\tp)\over 2\hbar}\phi_i^2\right),
\end{equation}
where
\begin{equation}
 \Omega_i (\tW;\tp) ={\displaystyle \alpha_{i}^{*}
 e^{-i\omega_i (t-\tp)}{dK_{Li}(\tW)\over id\tW}
 +\beta_{i}^{*} e^{i\omega_i (t-\tp)}{dQ_{Li}(\tW)\over id\tW}
 \over
 \alpha_{i}^{*}e^{-i\omega_i (t-\tp)}K_{Li}(\tW)
 +\beta_{i}^{*} e^{i\omega_i (t-\tp)}Q_{Li}(\tW)}.
\end{equation}
Here we assumed that the Bogoliubov coefficients in the
initial squeezed state are diagonal like
$\alpha_{ij}=\alpha_{i}\delta_{ij}$.
Since $\tilde \Phi^{\alpha\beta}_i(\bar X_L (\tW),\phi_i;\tp)$
completely determines the coherence factor as
\begin{equation}
 \prod_{i=1}^{D'} {\cal R}_{i}(t,t';\tp)
 =\prod_{i=1}^{D'} \int_{-\infty}^{\infty} d\phi_i
  \tilde \Phi^{\alpha\beta}_i(\bar X_L (\tW),\phi_i;\tp)
  \tilde \Phi^{\alpha\beta*}_i(\bar X_L (\tW'),\phi_i;\tp),
\end{equation}
in principle, we can calculate ${\cal R}_i (t,t';\tp)$,
but it is a formidable work to be done in practice.
So we consider a simple case in which
$\vert\beta_i/\alpha_i {\cal C}_i^2\vert<<1$.
However we do not assume
$\vert\beta_i e^{2\gamma_i}/\alpha_i{\cal C}_i^2\vert<<1$.
Roughly speaking, this means that we consider the
situation in which the initial excitation is not so large
but the excitation due to tunneling is not necessarily
larger than the initial excitation.
Then we obtain
\begin{eqnarray}
 i\Omega_i (\tW;\tp) &=&\left(
  {dK_{Li}(\tW)\over id\tW}K_{Li}^{-1}(\tW)
  -2i\omega_i^2
    {\beta^{*}_i e^{2i\omega_i (t-\tp)}\over
     \alpha^{*}_i K_{Li}^2(\tW)}\right)
     \left(1+O(\vert\beta_i/\alpha_i \vert)\right),
 \\
 {1\over i}{d\Omega_i (\tW;\tp)\over dt}
 &=&\left( m_{Ti}^2-\Omega_i^2 (\tW;\tp)- 4\omega_i^2
    {\beta^{*}_i e^{2i\omega_i (t-\tp)}\over
     \alpha^{*}_i K_{Li}^2(\tW)}\right)
     \left(1+O(\vert\beta_i/\alpha_i \vert)\right)
 \nonumber \\
 &=& \left(\left\{m_{Ti}^2
    -\left({dK_{Li}(\tW)\over id\tW}K_{Li}^{-1}(\tW)\right)^2
    \right\}- 4\omega_i(\omega_i +m_{Ti})
     {\beta^{*}_i e^{2i\omega_i (t-\tp)}\over
     \alpha^{*}_i K_{Li}^2(\tW)}\right)
     \left(1+O(\vert\beta_i/\alpha_i \vert)\right).
 \nonumber \\
 \label{Omegadott}
\end{eqnarray}
Under the assumption $\vert\beta_i/\alpha_i {\cal C}_i^2\vert<<1$,
$\Re(\Omega_i (\tW;\tp)$ reduces to
the same one given in (\ref{ReOmega}).
Noting that the absolute value of the first term
in the last line of Eq. (\ref{Omegadott}) is
$2m^2_{Ti}/(\cosh 2\gamma_i +\cos 2(m_{Ti}\tW+\varphi_i)),$
two extreme cases can be considered.

When $\vert\beta_i e^{2\gamma_i}/\alpha_i{\cal C}_i^2\vert<<1$,
the first term in the last line of Eq. (\ref{Omegadott})
dominates.
Therefore $1-{\cal R}_{i}(t,t';\tp)$ is not so different from
the value obtained in the previous case.
The difference is of $O(\vert\beta_i e^{2\gamma_i}
/\alpha_i{\cal C}_i^2\vert)$.
Therefore the degree of decoherence becomes the same order
as before.
This is expected because the excitations due to
initial condition is negligible compared with those
due to the tunneling.

On the other hand,
when $\vert\beta_i e^{2\gamma_i}/\alpha_i{\cal C}_i^2\vert>>1$,
The second term in the last line of Eq.~(\ref{Omegadott})
dominates.
In this case, ${\cal R}_{i}(t,t';\tp)$ is evaluated as
\begin{equation}
 {\cal R}_{i}(t,t';\tp)=1-{\omega_i^2 (\omega_i+m_{Ti})^2
  \vert\beta_i\vert^2
   \over m_{Ti}^2 \vert\alpha_i\vert^2
   \vert {\cal C}_i \vert^4}(t-t')^2+\cdots.
\end{equation}
Thus, comparing this with (\ref{vacR}), we find that
the degree of decoherence becomes larger than that in
the case initially in the {\it quasi-ground-state}.

So we conclude that the estimate of $1-{\cal R}_i(t,t';\tp)$ by
Eq.~(\ref{vacR}) gives the minimum degree of decoherence
in general.
Therefore, in the following discussion, we use the expression
given in (\ref{vacR}) for simplicity.

{}From the cosmological point of view,
the $O(4)$-symmetric vacuum bubble nucleation seems to be one of
the most interesting phenomena which relates to the quantum tunneling.
However, in that case, as every degree of freedom of the tunneling
field couples with each other, the analysis becomes very complicated.
Therefore, for the purpose to see to what extent we can justify the
identification of the two different flows of time in the field
theoretical problem,
we investigate a more tractable model such as the spatially
homogeneous decay model which was examined in Refs.
\cite{TaSaYa,Rubako}.

Let us consider the system which consists of two fields in
a finite volume $L^3$.
One is the tunneling field $\sigma$ and the other is the environment
$\phi$.
The Hamiltonian is given by
\begin{equation}
 H=H_{\sigma}+H_{\phi},
\end{equation}
where
\begin{eqnarray}
 H_{\sigma} & := & \int_{L^3} d^3\bx\left({1\over 2}
    p_{\sigma}^2+{1\over 2}\left(\nabla\sigma\right)^2
    +V(\sigma)\right)
\nonumber \\
 H_{\phi} & := & \int_{L^3} d^3\bx\left({1\over 2}
    p_{\phi}^2+{1\over 2}\left(\nabla\phi\right)^2
    +{1\over 2}m^2(\sigma)
    \phi^2\right),
\end{eqnarray}
where $p_{\sigma}$ and $p_{\phi}$ are the conjugate
momenta of $\sigma$
and $\phi$ respectively.
The potential $V(\sigma)$ has the form shown in Fig.~3.
If the spatial volume is infinite, the rate of tunneling
driven by the spatially homogeneous instanton,
$\sigma_0(\tau)$, is completely suppressed.
However, if a finite spatial volume is considered,
this tunneling process is relevant.

To apply the previous formalism to the present case,
we make the following correspondence,
\begin{equation}
 X(\tau)\rightarrow \sigma_0(\tau), \quad
  \phi_i(\tau)\rightarrow \phi_k(\tau)
 :={1\over(2\pi)^{3/2}}\int_{L^3}
 d^3\bx~e^{-i\bk\bx}\phi(\bx;\tau).
\end{equation}
Hereafter, we neglect the existence of the fluctuation
degrees of freedom of the $\sigma$ field itself.
Further, for simplicity, we restrict the $\sigma$
dependence of the $\phi$-field mass $m^2(\sigma)$
to be that given by a step function;
\begin{equation}
 m^2(\sigma)=\left\{\begin{array}{ll}
   \mbox{$ m^2_{-}$} & \mbox{$(\sigma<\tilde\sigma),$} \\
   \mbox{$ m^2_{+}$} & \mbox{$(\sigma>\tilde\sigma).$}
\end{array}
   \right.
\end{equation}
We assume that $\sigma_F<\tilde\sigma<\sigma_T$ and introduce the
WKB time
$\tilde\tau(<0)$ at which $\sigma_0(\tilde\tau)=\tilde\sigma$.

Under this circumstance, the unnormalized negative frequency function
$K_{L\bk}^{*}(\tW)$
specifying the state after tunneling is given by
\begin{equation}
 K_{L\bk}=A_{\bk} e^{i\omega_{+}\tW}+B_{\bk} e^{-i\omega_{+}\tW},
\end{equation}
where $\omega_{\pm}:=\sqrt{\bk^2+m^2_{\pm}}$ and
\begin{eqnarray}
 A_{\bk} & = & {1\over 2\omega_{+}}(\omega_{+}+\omega_{-})
    e^{-(\omega_{+}-\omega_{-})\tilde\tau},
 \nonumber \\
 B_{\bk} & = & {1\over 2\omega_{+}}(\omega_{+}-\omega_{-})
    e^{(\omega_{+}+\omega_{-})\tilde\tau}.
\end{eqnarray}
Therefore we can read
\begin{equation}
 e^{2\gamma_{\bk}+i\varphi_{\bk}}={B_{\bk}\over A_{\bk}}
 ={\omega_{+}-\omega_{-}\over\omega_{+}+\omega_{-}}e^{2\omega_{+}\tilde\tau}.
\end{equation}
Integrating over all Fourier components, we obtain
\begin{equation}
 \tilde\rho(\bar X_L(\tW),\bar X_L (\tW'))
 =\Theta(\bar X_L(\tW))\Theta^{*}(\bar X_L(\tW'))\times
   \left[1-{L^3\over(2\pi)^3}\int d^3 \bk
   {4\omega_{+}^2\over\sinh^2 2\gamma_{\bk}}(\tW-\tW')^2 +\cdots\right],
\end{equation}
The second term in the square bracket is evaluated to give
\begin{equation}
 \sim (\tW-\tW')^2\times\left\{\begin{array}{lll}
     \mbox{$\displaystyle
     L^3(\Delta m^2)^2\vert\tilde\tau\vert^{-1}$}
     & \hbox{\rm for} &
     \mbox{$\displaystyle \vert\tilde\tau\vert << m_{+}^{-1},$}
     \\
     \mbox{$\displaystyle
     L^3 m_{+}^3 m_{-}^{-2}(\Delta m^2)^2 $}
     & \hbox{\rm for} &
     \mbox{$\displaystyle \vert\tilde\tau\vert >> m_{+}^{-1}.$}
\end{array}
  \right.
\end{equation}
Thus we conclude that if the volume is large enough compared to the
the inverse mass scale, namely, the Compton length of the $\phi$ field,
the two states with the difference of $\tW$ larger
than the Compton time scale of the $\phi$ field loses their coherence
after tunneling.

Although it is difficult to extract some information about the
$O(4)$-symmetric bubble nucleation from this simple toy model,
we expect when the nucleated bubble becomes large enough
compared to the wall thickness, which becomes the same order
of the Compton length of the tunneling field itself,
the WKB trajectory of the wall becomes classical
and the WKB time can be identified with the external time
with error less than the scale of the wall thickness.

The laboratory experiments in a situation when
this identification is not allowed would be very interesting topic
in the future.

\section{Conclusion}

We considered a problem concerning the quantum tunneling with
coupling to environmental degrees of freedom.
In the previous work, the situation in which the initial
quantum state is in an energy eigen state was considered.
Here, considering an extension to the case in which
the quantum state is in a squeezed state, we found that
there is a problem of identification of two different
flows of time, {\it i.e.}, the WKB time and the external time.
The WKB time is just a parameter which parametrizes the
configuration space along the classical trajectory.
We pointed out that this identification plays a crucial
role to give a simple interpretation and understanding of
the quantum state after tunneling, especially for the
tunneling from a squeezed state.
Long enough after we set the initial state in the false vacuum,
the wave function may develop into the state of a superposition
of the wave packets which represent the tunneling occurred at
different moments.
We called them sectors.
In each sector,
we can identify the two different flows of time.
Therefore,
if we can think of this state not as a quantum superposition
but as a statistical ensemble of different states, in other words,
the quantum coherence between different sectors is lost,
the identification will be justified.
Thus we considered the loss of the quantum coherence
in the tunneling situation as a mechanism of this identification.
The different sectors are parametrized by the different WKB times.
The larger the difference of the WKB times,
the less the sectors will be coherent.
Therefore there is a typical scale of the difference of WKB times
where quantum coherence is lost.
So we estimated this time scale of decoherence using a toy model
of a spatially homogeneous decay, and
we obtained that the time scale of decoherence becomes shorter
than the Compton time scale of the field coupling to the
tunneling field unless the coupling is extremely weak.

\vskip 0.1in
\centerline{ACKNOWLEDGMENTS}
\vskip 0.05in
I would like to express my thanks to Prof. Misao Sasaki
and Dr. Kazuhiro Yamamoto for useful discussions and comments.
I also thank Prof. Humitaka Sato for continuous
encouragement.
This work was supported by Monbusho Grant-in-Aid for
Scientific Research No. 2010.

\vskip2cm
\begin{figure}
\caption{The potential form of the tunneling degree of freedom,
where $X_F$ represents the values of
$X$ in the false vacuum.
}
\end{figure}
\begin{figure}
\caption{A path of integration on the complex plane of time.
}
\end{figure}
\begin{figure}
\caption{The potential form of the tunneling field.}
\end{figure}


\begin{references}
\bibitem{Recent}
 M. Sasaki, T. Tanaka and K. Yamamoto, in {\it preparation}.

\bibitem{Inf}
 D.~La and P.~J.~Steinhardt, Phys. Rev. Lett. {\bf 62} (1989) 376;
 P.~J.~Steinhardt and F.~S.~Accetta, Phys. Rev. Lett. {\bf 64}
 (1990) 2740;
 F.~C.~Adams and K.~Freese, Phys. Rev. {\bf D43} (1991) 353;

\bibitem{TaSaYa}
 T. Tanaka, M. Sasaki and K. Yamamoto,
Phys. Rev. {\bf D49} (1994) 1039.

\bibitem{VeGeSa}
 H.J. de Vega, J.L. Gervais and B. Sakita,
 Nucl. Phys. {\bf B139} (1978) 20;
 H.J. de Vega, J.L. Gervais and B. Sakita,
 Nucl. Phys. {\bf B143} (1978) 125.

\bibitem{Vega}
 H.J. de Vega, J.L. Gervais and B. Sakita,
 Phys. Rev. {\bf D19} (1979) 604.

\bibitem{SaTaYY}
  M. Sasaki, T. Tanaka, K. Yamamoto and J. Yokoyama,
Prog. Theor. Phys. {\bf 90} (1993) 1019;
K. Yamamoto, T. Tanaka, and M. Sasaki, in {\it preparation}

\bibitem{Rubako}
  V.A.~Rubakov, Nucl. Phys. {\bf B245} (1984) 481.

\bibitem{VacVil}
  T.~Vachaspati and A.~Vilenkin, Phys. Rev. {\bf D43} (1991) 3846.

\bibitem{Deco}
 See, e.g., W. H. Zurek, Prog. Theor. Phys. {\bf 89} (1993) 281,
 and references therein;


\bibitem{Yamamoto}
K.~Yamamoto, Prog. Theor. Phys. {\bf 91} (1994) 437.

\bibitem{VacVilC}
T.~Vachaspati and A.~Vilenkin, Phys. Rev. {\bf D37} (1988) 898.

\bibitem{EffGra}
 T. Tanaka and M. Sasaki, Kyoto university preprint KUNS1267,
 Phys. Rev. {\bf D} {\it to be published}.

\bibitem{strate}
 E. Joos and H. D. Zeh, Z. Phys. {\bf B59} (1985), 223;

\end{references}
\end{document}